\documentclass[twocolumn,english,aps,prb,floatfix,preprintnumbers,showpacs,amsfonts,amssymb,superscriptaddress]{revtex4}
\usepackage{ae,aecompl}
\usepackage[T1]{fontenc}
\usepackage[latin1]{inputenc}
\usepackage{amsmath}
\usepackage{graphicx}
\usepackage{amssymb}

\makeatletter

\providecommand{\tabularnewline}{\\}

\@ifundefined{textcolor}{}
{%
 \definecolor{BLACK}{gray}{0}
 \definecolor{WHITE}{gray}{1}
 \definecolor{RED}{rgb}{1,0,0}
 \definecolor{GREEN}{rgb}{0,1,0}
 \definecolor{BLUE}{rgb}{0,0,1}
 \definecolor{CYAN}{cmyk}{1,0,0,0}
 \definecolor{MAGENTA}{cmyk}{0,1,0,0}
 \definecolor{YELLOW}{cmyk}{0,0,1,0}
 }

\usepackage{amscd}\usepackage{bm}

\makeatother

\usepackage{babel}

\makeatother

\usepackage{babel}

\begin{document}

\title{Rényi Entropy and Parity Oscillations of the Anisotropic Spin-\emph{s}
Heisenberg Chains in a Magnetic Field}

\author{J.~C.~Xavier}

\affiliation{Instituto de Física, Universidade Federal de Uberlândia, Caixa Postal
593, 38400-902 Uberlândia, MG, Brazil }

\affiliation{Instituto de Física de São Carlos, Universidade de São Paulo, Caixa
Postal 369, 13560-970 São Carlos, SP, Brazil}

\author{F.~C.~Alcaraz}

\affiliation{Instituto de Física de São Carlos, Universidade de São Paulo, Caixa
Postal 369, 13560-970 São Carlos, SP, Brazil}

\date{\today{}}
\begin{abstract}
Using the density matrix renormalization group, we investigate the
Rényi entropy of the anisotropic spin-$s$ Heisenberg chains in a
$z$-magnetic field. We considered the half-odd-integer spin-$s$
chains, with $s=1/2,3/2$ and $5/2$, and periodic and open boundary
conditions. In the case of the spin-1/2 chain we were able to obtain
accurate estimates of the new parity exponents $p_{\alpha}^{(p)}$
and $p_{\alpha}^{(o)}$ that gives the power-law decay of the oscillations
of the $\alpha-$Rényi entropy for periodic and open boundary conditions,
respectively. We confirm the relations of these exponents with the
Luttinger parameter $K$, as proposed by Calabrese \emph{et al. }{[}Phys.
Rev. Lett. \textbf{{104}}, 095701 (2010){]}. Moreover, the predicted
periodicity of the oscillating term was also observed for some non-zero
values of the magnetization $m$. We show that for $s>1/2$ the amplitudes
of the oscillations are quite small, and get accurate estimates of
$p_{\alpha}^{(p)}$ and $p_{\alpha}^{(o)}$ become a challenge. Although
our estimates of the new universal exponents $p_{\alpha}^{(p)}$ and
$p_{\alpha}^{(o)}$ for the spin-3/2 chain are not so accurate, they
are consistent with the theoretical predictions. 
\end{abstract}


\maketitle

\section{INTRODUCTION}

The observation that entanglement may play an important role at a
quantum phase transition has motivated many studies on the characterization
of the critical phenomena by using quantum information concepts.\cite{naturefazio,praosborne,cvidal,revfazio,prlkorepin,cardyentan,chicosarandy}
Quantum spin chains have been proven as useful laboratories to investigate
the interconnection of entanglement and quantum criticality.\cite{naturefazio,praosborne,cvidal,revfazio,prlkorepin,cardyentan,chicosarandy}
Although does not exist yet an universal measure that quantifies the
entanglement,\cite{reviewriera} the von Neumann entropy and the Rényi
entropies are the most commonly used measures since they are sensitive
to the long-distance quantum correlations of critical systems. 

In this paper, we study the Rényi entropy in the critical region of
the anisotropic spin-$s$ Heisenberg models for $s=1/2$, 3/2, and
5/2. Consider an one-dimensional system of size $L$ and composed
by two subsystems $A$ and $B$ of sizes $l$ and $L-l,$ respectively.
The Rényi entropy is defined as

\begin{equation}
S_{\alpha}(L,l)=\frac{1}{1-\alpha}\ln Tr(\rho_{A}^{\alpha}),\label{eq:2b}\end{equation}
 where $\rho_{A}$ is the reduced density matrix of the subsystem
$A$. The von Neumann entropy is given by the limiting case $\alpha=1$.

In the last few years a great effort has been made to understand the
asymptotic behavior of $S_{\alpha}(L,l)$. It is expected that the
ground state of critical one-dimensional systems gives a Rényi entropy
that behaves as

\begin{equation}
S_{\alpha}(L,l)=S_{\alpha}^{CFT}(L,l)+S_{\alpha}^{osc}(L,l).\label{eq:3}\end{equation}
 The first term, in this equation, is the conformal field theory (CFT)
prediction in the scaling regime ($L\gg l\gg1)$, and is given by
\citealp{cold,cardyentan,entroreviewcalabrese,cvidal,affleckboundary}

\begin{equation}
S_{\alpha}^{CFT}=\frac{c}{6}\left(1+\frac{1}{\alpha}\right)\ln\left[\left(\frac{L}{\pi}\sin\left(\frac{\pi l}{L}\right)\right)\right]+c_{1}^{(p)},\label{eq:3b}\end{equation}
for periodic boundary conditions (PBC), and

\begin{equation}
S_{\alpha}^{CFT}=\frac{c}{12}\left(1+\frac{1}{\alpha}\right)\ln\left[\frac{4(L+1)}{\pi}\sin\left(\frac{\pi(2l+1)}{2(L+1)}\right)\right]+c_{1}^{(o)},\label{eq:3OBC}\end{equation}
for open boundary conditions (OBC). $c$ is the central charge, and
$c_{1}^{(p)}$ and $c_{1}^{(o)}$ are non-universal constants. The
expression (\ref{eq:3OBC}) for OBC has a small modification that
is absorbed in the constant $c_{1}^{(o)}$, as compared with the standard
CFT expression.\cite{cardyentan,cold,entroreviewcalabrese,affleckboundary}
We chosed this expression since, as shown in Ref. \onlinecite{calabreseOBC},
in the case of the XX model, it is accurate up to order $(1/l)$ (see
also Ref. \onlinecite{affleckosc2}). It is interesting to mention
that a generalization of the above equations for the excited states
was proposed recently.\citealp{chicoentropy}

The second term in Eq. (\ref{eq:3}) seems to origin in the strong
antiferromagnet correlations, as first argued by Laflorencie \emph{et
al}. in the case of the spin-1/2 Heisenberg chain with OBC.\citealp{entropyaffleckosc}
In open chains, since translation invariance is broken, the energy
density as a function of the site $l$ also shows a similar decaying
alternating term $E^{osc}$. In Ref. \onlinecite{entropyaffleckosc}
by using bosonization techniques this oscillating term was calculated
and compared with numerical evaluations of the entropy suggesting
that $S_{1}^{osc}(l)\sim E^{osc}\sim\left[\frac{L}{\pi}\sin\left(\frac{\pi l}{L}\right)\right]^{K}$,
where $K$ is the Luttinger parameter. In fact, these strong oscillations
for the open chains were observed by several authors. \citealp{entropyaffleckosc,zhouosc,peterentropy,Kollath,entrosoc2,roax,affleckosc2,sierra-cirac,xavierentanglement,entropyosc,xxPBCh,calabreseOBC,calabreserandom}

More recently, Calabrese \emph{et al}. in Ref. \onlinecite{entropyosc}
(see also Ref. \onlinecite{xxPBCh}) investigate the anisotropic
spin-1/2 Heisenberg chain with periodic boundary conditions and verified
that even in this case those oscillations are still present if the
Rényi index $\alpha>$1. For the XX chain in a magnetic field with
PBC/OBC, Calabrese and collaborators obtained exactly $S_{\alpha}^{osc}$,
and observed an universal behavior of this oscillating term.\citealp{entropyosc,xxPBCh,calabreseOBC}
Based on these exact results and on numerical calculations of the
spin-1/2 XXZ chain with PBC at zero magnetic field, Calabrese and
collaborators conjectured that $S_{\alpha}^{osc}$ has the following
universal behavior\citealp{entropyosc,xxPBCh,calabreseOBC}(see also
Ref. \onlinecite{entropyaffleckosc}) \begin{equation}
S_{\alpha}^{osc}=\frac{g_{\alpha}^{(p)}}{L^{p_{\alpha}^{(p)}}}\cos\left(2lk_{F}\right)\left|\sin\left(\pi\frac{l}{L}\right)\right|^{-p_{\alpha}^{(p)}},\label{eq:3c}\end{equation}
for PBC and

\begin{equation}
S_{\alpha}^{osc}=\frac{g_{\alpha}^{(o)}}{L^{p_{\alpha}^{(o)}}}\sin\left[(2l+1)k_{F}^{(o)}\right]\left|\sin\left[\frac{\pi(2l+1)}{2(L+1)}\right]\right|^{-p_{\alpha}^{(o)}},\label{eq:3cOBC}\end{equation}
for OBC. The constants $g_{\alpha}^{(p)}$ and $g_{\alpha}^{(o)}$
are non-universal and the critical exponents governing the decaying
of the amplitudes of the oscillations are $p_{\alpha}^{(p)}=2p_{\alpha}^{(o)}=\frac{2K}{\alpha}$.
In the PBC case {[}Eq. (\ref{eq:3c}){]} the period of the oscillations
depends on the Fermi momentum $k_{F}$, while in the OBC case {[}Eq.
(\ref{eq:3cOBC}){]} it depends on $k_{F}^{(o)}=\frac{L}{L+1}k_{F}+\frac{\pi}{2(L+1)}$.

The origin of the exponents $p_{\alpha}^{(p)}$ and $p_{\alpha}^{(o)}$,
as observed by Cardy and Calabrese in Ref. \onlinecite{cardyosc},
are the conical spacial-time singularities produced in the conformal
mapping used to describe the reduced density matrix $\rho_{A}$ in
the CFT. There are two important ingredients in the oscillatory behavior
of Eqs. (\ref{eq:3c}) and (\ref{eq:3cOBC}): the non-universal constants
$g_{\alpha}^{(p)}$ and $g_{\alpha}^{(o)}$ and the Luttinger parameter
$K$ (that gives the exponents $p_{\alpha}^{(p)}$ and $p_{\alpha}^{(o)}$).
Unfortunately, there is no prediction for $g_{\alpha}^{(p)}$ and
$g_{\alpha}^{(o)}$ and as we are going to verify, in the case of
the spin-$s$ anisotropic Heisenberg model, they decrease dramatically
as we increase the spin size $s$. 

Our aim in the present paper is to verify the above conjectures more
extensively for the spin-1/2 XXZ, as well for the critical regions
of the spin $s>1/2$ anisotropic Heisenberg model in the presence
of an external $z$-magnetic field. The inclusion of the magnetic
field is interesting because the magnetization and the Luttinger parameter
$K$ depend on its value. We have then more possibilities to verify
the conjectures (\ref{eq:3c}) and (\ref{eq:3cOBC}). The parameter
$K$ in the case of the spin-1/2 XXZ chain can be calculated exactly
from the Bethe ansatz solution of the model. For $s>1/2$ the model
is not exactly integrable but this parameter can be calculated numerically
by exploring the consequences of the conformal invariance of the quantum
chain in the bulk limit. In the following we present some important
relations that will be used to evaluate the Luttinger liquid parameter
$K$.

The ground state energy of a system of size $L$, as $L\rightarrow\infty$,
behaves as\citealp{anomaly1,anomaly2}

\begin{equation}
\frac{E_{0}}{L}=e_{\infty}+\frac{f_{\infty}}{L}-\frac{v_{s}\pi c}{\delta6L^{2}}+o(L^{-2}),\label{eq:1}\end{equation}
 where $\delta=1$($\delta=4$) for the system with periodic (open)
boundary condition, $v_{s}$ is the sound velocity, $e_{\infty}$
is the bulk ground state energy per site, and $f_{\infty}$ is the
surface free energy, that vanishes for the systems with PBC.

The mass gap amplitudes of the finite-size corrections of the higher
energy states, for a system with periodic (open) boundary conditions,
are related to the anomalous dimensions $x_{bulk}^{\beta}$ (surface
exponents $x_{s}^{\beta}$). In the periodic case there are, for each
primary operator $O_{\beta}$ ($\beta=1,2,...)$ in the CFT, a tower
of states $E_{j,j'}^{\beta}(L)$ in the spectrum of the Hamiltonian
with asymptotic behavior \citealp{dimensions}

\begin{equation}
E_{j,j'}^{\beta}(L)-E_{0}(L)=\frac{2\pi v_{s}}{L}(x_{bulk}^{\beta}+j+j')+o(L^{-1}),\label{eq:2}\end{equation}
 where $j,j'=0,1,2,...$. For the chains with OBC the tower of states
have energies\citealp{surfexp}

\begin{equation}
E_{j}^{\beta}(L)-E_{0}(L)=\frac{\pi v_{s}}{L}(x_{s}^{\beta}+j)+o(L^{-1}),\label{eq:2x}\end{equation}
 with $j=0,1,2,...$.

For models described by a Luttinger liquid CFT, which are the present
cases, the Luttinger liquid parameter $K$ is given by $K=\frac{1}{4x_{p}}$,
where $x_{p}$ is the lowest anomalous dimension obtained by using
in Eq. (\ref{eq:2}) the lowest excited state in the sector whose
magnetization is increased by one unit with respect to that of the
ground state.

\section{THE MODEL}

We consider the anisotropic spin-$s$ Heisenberg chain, also known
as the spin-$s$ XXZ chain, in the presence of a magnetic field $h$
with Hamiltonian given by

\begin{equation}
H=\sum_{j}\left(s_{j}^{x}s_{j+1}^{x}+s_{j}^{y}s_{j+1}^{y}+\Delta s_{j}^{z}s_{j+1}^{z}\right)-h\sum_{j}s_{j}^{z},\label{eq:4}\end{equation}
 were $\vec{s}_{j}=(s_{j}^{x},s_{j}^{y},s_{j}^{z})$ are the spin-$s$
$SU(2)$ operators and $\Delta=\cos\gamma$ is the anisotropy.

We investigate the above model, using the density matrix renormalization
group (DMRG)\citealp{white} method with OBC and PBC, keeping up to
$\tilde{m}$=4000 states per block in the final sweep. We have done
$\sim6-11$ sweeps, and the discarded weight was typically $10^{-7}-10^{-12}$
at that final sweep. In our DMRG procedure the center blocks are composed
of (2$s$+1) states.

Let us first present some known results \emph{in the absence of the
magnetic field}, i.e, $h=0$. It is well known that this model at
the isotropic point $\Delta=1$ or $\gamma=0$ is gapless (gapful)
for half-odd-integer (integer) spins. \citealp{haldaneconj1,haldaneconj2}
The anisotropic chains are critical and conformal invariant for $-1<\Delta\le1$
with central charge $c=1$ for half-odd-integer spins. \citealp{schulz3,spinsaffleck,spinschico,hallbergetal}
On the other hand, in the case of integer spins a critical phase appears
for $-1\le\Delta\le\Delta_{c}(s)$, where $\Delta_{c}(s)<1$ is a
critical anisotropy.\citealp{spinschico}

The spin-1/2 XXZ chain is exactly soluble\citealp{chico1,chico2}
and for this reason some exact results are known on its critical region
$-1\leq\Delta=\cos\gamma\leq1$. In particular, the anomalous dimension
(surface exponent) associated to the lowest eigenenergy, in the sector
with total spin $z$-component $S_{T}^{z}=\sum_{j}s_{j}^{z}=1$, is
given by $x_{p}=\frac{\pi-\gamma}{2\pi}$ ($x_{s}=2x_{p}$) and the
sound velocity is $v_{s}=\frac{\pi\sin\gamma}{2\gamma}$. \citealp{chico1,chico2,xxzhamer,xxxWoynarovich,lutherpeschel2}
The exact solution of the spin-1/2 chain has also been explored in
the context of the entanglement calculations. \citealp{satoxxz1,banchi,satoxxz2,alba,nienhuiscalabrese}
Although exactly integrable, in this context, only some few issues
were explored in the spin-1/2 XXZ chain. This is due to the difficulty
in extracting analytically results from its exact solution.

For $h\ne0$ the model is in a critical and conformal invariant phase
for $h_{c}<|h|<1+\Delta$. The critical field $h_{c}=0$ for $|\Delta|\le1$,
and for $\Delta>$1, $h_{c}=h_{c}(\Delta)$ depends continuously on
$\Delta$. In this phase the model is described by a Luttinger liquid
phase whose parameter $K=\frac{1}{4x_{p}}$ depends on the values
of the magnetization per site $m=m(h)=\frac{S_{T}^{Z}}{L}$ of the
ground state, and the anisotropy $\Delta$. The exact solution of
the model allows us to obtain the exponents $x_{p}=x_{p}(m)$. This
is done by solving a set of non-linear integral equations that, for
the sake of brevity we refer to Ref. \onlinecite{xxzhWoy}. It is
expected that for the half-odd spins $s$, with $s>$1/2, similar
phases emerge when a magnetic field is applied.

\section{RESULTS}

As mentioned earlier, we need to calculate the anomalous dimension
$x_{p}$ in order to verify if the oscillating term of the Rényi entropies
{[}Eqs. (\ref{eq:3c}) and (\ref{eq:3cOBC}){]} decays with the new
exponents $p_{\alpha}^{(p)}=2p_{\alpha}^{(o)}=\frac{2K}{\alpha}$.
For the spin-1/2 case and $h=0$, we know that $x_{p}(m=0)=\frac{\pi-\gamma}{2\pi}$.
For $h\ne0$ we can still determine $x_{p}(m)$ exactly by solving
numerically a set of non-linear integral equations given in Ref. \onlinecite{xxzhWoy}.
On the other hand, for $s>1/2$ there are no exact results for $x_{p}$.

In order to verify the dependence of the exponents $p_{\alpha}^{(p)}$
and $p_{\alpha}^{(o)}$ with the Luttinger parameter $K$ (or $x_{p})$,
we need to use independent estimates for them. The independent estimates
can be obtained from the mass gap of the eigenspectrum and from the
$\alpha-$Rényi entropies. We are going to calculate these quantities
using the DMRG technique. The estimate of $x_{p}$ will be obtained
by following basically the same procedure used by one of us in Ref.
\onlinecite{xavierentanglement}. We consider the Hamiltonian defined
in Eq. (\ref{eq:4}) with PBC to determine the anomalous dimension
$x_{p}$. The value of $x_{p}(m)$ is obtained from the limit $L\rightarrow\infty$
of the finite-size sequences

\begin{equation}
x_{p}(m,L)=\frac{L\left[E(L,S_{T}^{z}+1)-E(L,S_{T}^{z})\right]}{2\pi v_{s}},\label{eq:xp}\end{equation}
where as before $S_{T}^{z}=\sum_{j}s_{j}^{z}$. 

We estimated the sound velocity $v_{s}$ using Eq. (\ref{eq:1}) with
$c=1$. We also assume that $x_{p}(m,L)$ behaves asymptoticaly as

\begin{equation}
x_{p}(m,L)=x_{p}(m)+a_{1}/L^{\omega}+a_{2}/L^{2},\label{eq:xpb}\end{equation}
 where $\omega=\frac{2}{x_{p}}-4$. These corrections are expected
from the finite-size perturbation of the critical models (see Ref.
\onlinecite{chico1}). We use a simple fit procedure to obtain $x_{p}(m)$
by considering typically system of sizes $L=16-96$.

\begin{table}
\begin{tabular}{ccccc}
\hline 
$\Delta$  & $m=0$  & $m=1/6$  & $m=1/4$  & $m=3/10$\tabularnewline
\hline 
$0$  & 0.2500  & 0.2499  & 0.2498  & 0.2498\tabularnewline
 & (\emph{0.25})  & (\emph{0.25})  & (\emph{0.25})  & (\emph{0.25})\tabularnewline
$0.5$  & 0.3333  & 0.3111  & 0.2956  & 0.2863\tabularnewline
 & (\emph{0.3333})  & (\emph{0.3118})  & (\emph{0.2959})  & (\emph{0.2863})\tabularnewline
$\sqrt{2}/2$  & 0.3752  & 0.3319  & 0.3106  & 0.2963\tabularnewline
 & (\emph{0.375})  & (\emph{0.3334} )  & ( \emph{0.3101})  & (\emph{0.2967})\tabularnewline
$0.9980$  & 0.464  & 0.358  & 0.327  & 0.309\tabularnewline
 & (\emph{0.49})  & (\emph{0.3599})  & (\emph{0.3264})  & (\emph{0.3086})\tabularnewline
$2.0$  & ----  & 0.422  & 0.363  & ----\tabularnewline
 & ----  & (\emph{0.4233})  & (\emph{0.3626})  & ----\tabularnewline
 &  &  &  & \tabularnewline
\end{tabular}

\caption{The anomalous dimension $x_{p}$ for the spin-1/2 XXZ chain with PBC
and some values of the magnetization $m$ and anisotropy $\Delta$.
The values in between the parentheses are the exact ones (see text). }

\end{table}

\subsection{s=1/2}

As a benchmark test, we consider first the spin-1/2 XXZ chain. In
Table I, we present the estimated values of $x_{p}$, for some values
of magnetization $m$ and anisotropy $\Delta$, obtained from Eqs.
(\ref{eq:xp}) and (\ref{eq:xpb}). We see on this table a clear agreement
between our numerical results, obtained via DMRG, and the exact values
shown in between the parentheses. We thus see that this procedure
gives accurate estimates of $x_{p}$ (similar results was found in
Ref. \onlinecite{xavierentanglement} for $m=0$). As mentioned
before, for $\Delta>1$ the model is still critical in the region
where the magnetic field produces a non-zero magnetization ($0<m<1/2$).
\citealp{xxzh1,xxzh2,xxzh3,xxzh4} Unfortunately, for some values
of $\Delta$ in this region, we observed that the DMRG is not stable.
In particular, for systems with OBC and $m\gtrsim3/10$ we observed
that the DMRG is not stable for any $\Delta$. For this reason, we
were able to estimate $x_{p}$ only for $\Delta\lesssim2$ and $m<3/10$.

\begin{figure}
\begin{centering}
\includegraphics[scale=0.27]{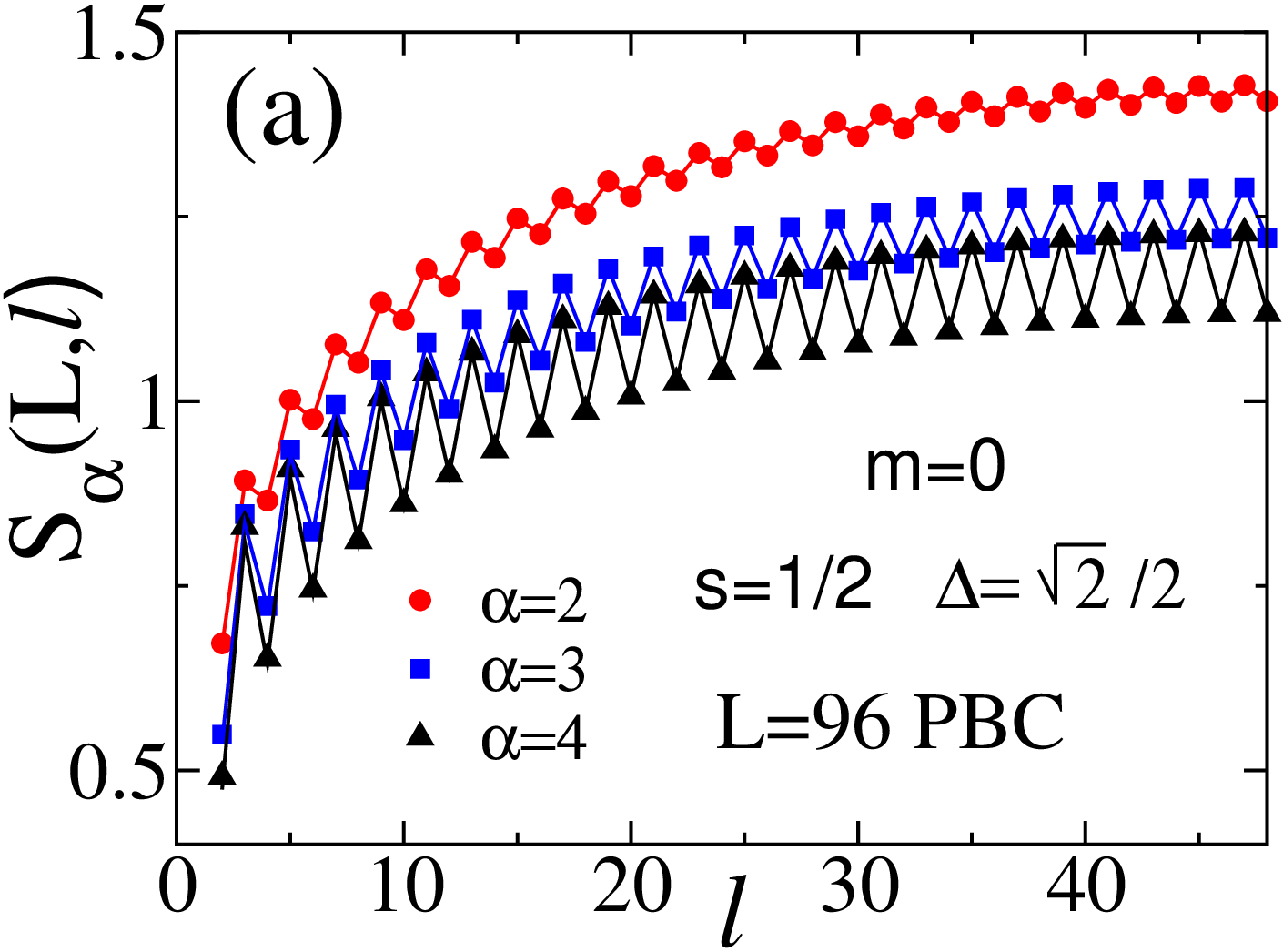} 
\par\end{centering}

\begin{centering}
\includegraphics[scale=0.27]{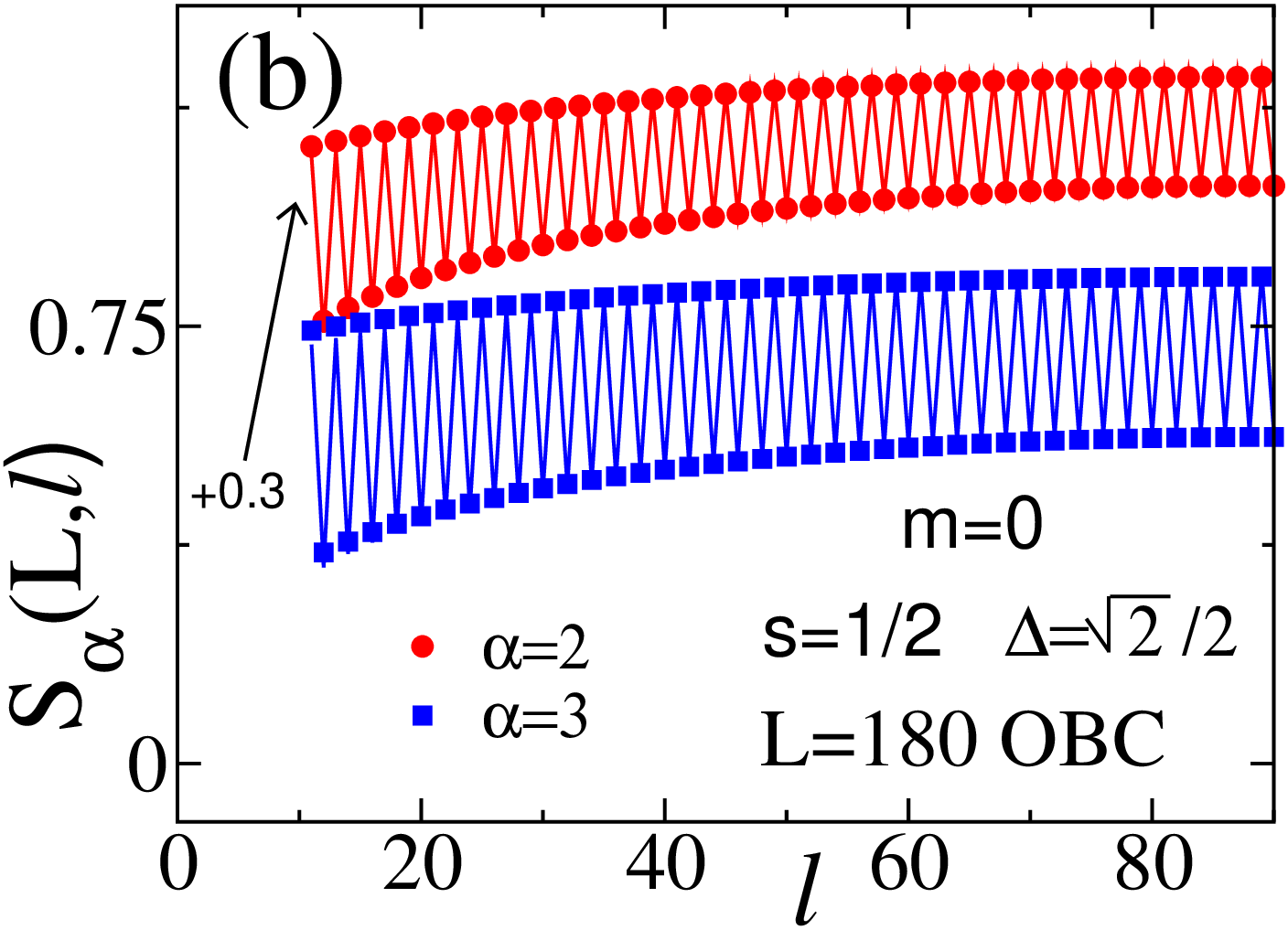} 
\par\end{centering}

\begin{centering}
\includegraphics[scale=0.27]{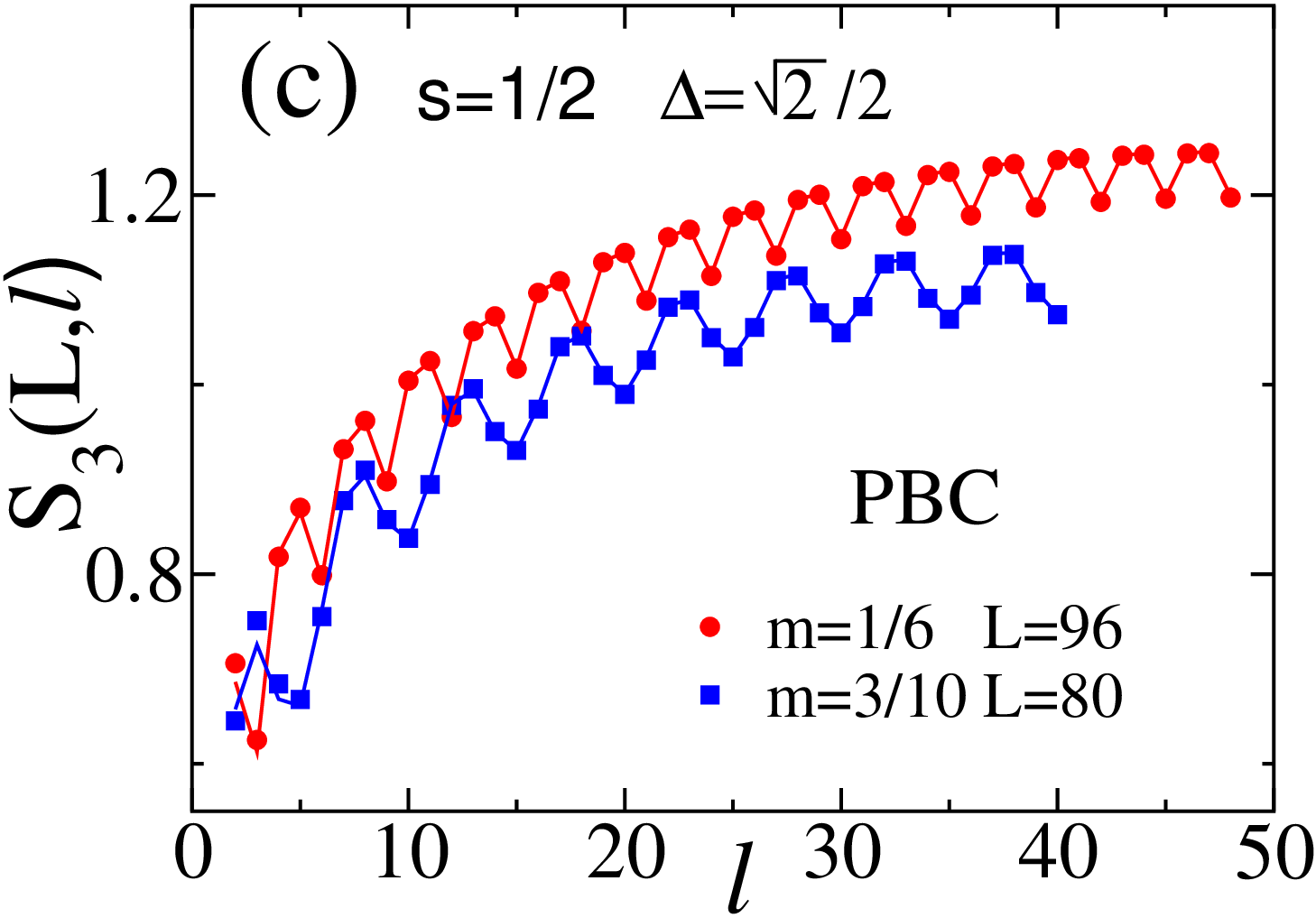} 
\par\end{centering}

\begin{centering}
\includegraphics[scale=0.27]{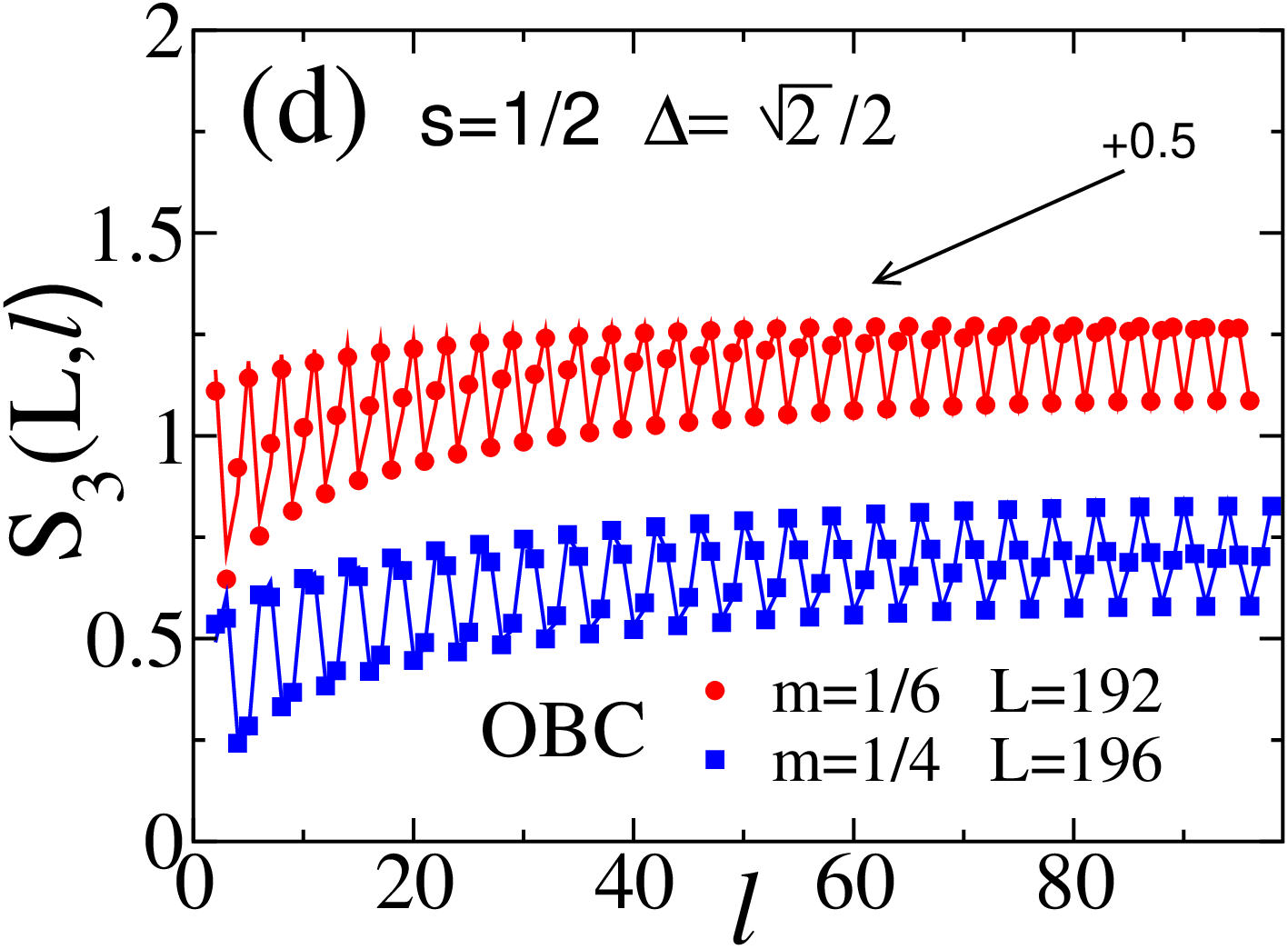} 
\par\end{centering}

\caption{\label{fig1} (Color online). The Rényi entropy $S_{\alpha}(L,l)$
as a function of $l$ for the spin-1/2 XXZ chain with $\Delta=\sqrt{2}/2$.
(a) $L=96$, PBC, $m=0$ and three values of $\alpha$ (see legend).
(b) $L=180$, OBC, $m=0$, and two values of $\alpha$ (see legend).
We added 0.3 in the values of the entropy for $\alpha=2$ in order
to see both data in the same figure. (c) $L=96$ with $m=1/6$, and
$L=80$ with $m=3/10$. In both bases $\alpha=3$ and PBC (d) $L=192$
with $m=1/6$, and $L=196$ with $m=3/10$. In both bases $\alpha=3$
and OBC.}

\end{figure}

Now, let us estimate the exponents $p_{\alpha}^{(p)}$ and $p_{\alpha}^{(o)}$.
Recently Calabrese \emph{et al.} in Ref. \onlinecite{entropyosc}
(see also Ref. \onlinecite{xxPBCh}) calculated exactly these exponents
for the spin-1/2 XX chain ($\Delta=0$) with PBC. They also verified
indirectly, through the DMRG algorithm, that $p_{\alpha}^{(p)}=\frac{1}{2x_{p}\alpha}$
for $\Delta\ne0$ and $h=0$. Fagotti and Calabrese in Ref. \onlinecite{calabreseOBC}
also obtained $p_{\alpha}^{o}$ exactly for the spin-1/2 XX chain
in a magnetic field with OBC. We also include in our analysis the
chains with OBC to show, indeed, that the conjecture holds for $\Delta\ne0$.
Here, instead of just verifying that the oscillating term $S_{\alpha}^{osc}$
is consistent with a universal decay mediated by the exponents $p_{\alpha}^{(p)}$
and $p_{\alpha}^{(o)}$, as done in Ref. \onlinecite{entropyosc},
we are going to estimate \emph{directly} these exponents. We obtain
our estimates from the direct fit of the numerical data to the functions
(\ref{eq:3c}) and (\ref{eq:3cOBC}).

In Figs. 1 and 2, we present the Rényi entropy $S_{\alpha}(L,l)$
as a function of $l$ for the anisotropic spin-1/2 Heisenberg chains
with PBC/OBC and for some values of $m$. The symbols (circle, squares,
etc) are the numerical data and the solid lines {[}with the exception
of Fig. 2(a){]} connect the fitted points using Eq. (\ref{eq:3})
with $c=1$. We show $S_{\alpha}(L,l)$ only for $l\le L/2$ since
$S_{\alpha}(L,L-l)=S_{\alpha}(L,l)$. As shown in Fig. 1(a), the amplitudes
of the oscillations increase with the value of $\alpha$. Similar
results were also observed in Ref. \onlinecite{entropyosc}. It
is important to stress that if $\alpha>$1 we can only get a reasonable
fit of the numerical data by considering the oscillating term $S_{\alpha}^{osc}$,
in addition to the standard term $S_{\alpha}^{CFT}$ predicted by
CFT.

\begin{figure}
\begin{centering}
\includegraphics[scale=0.27]{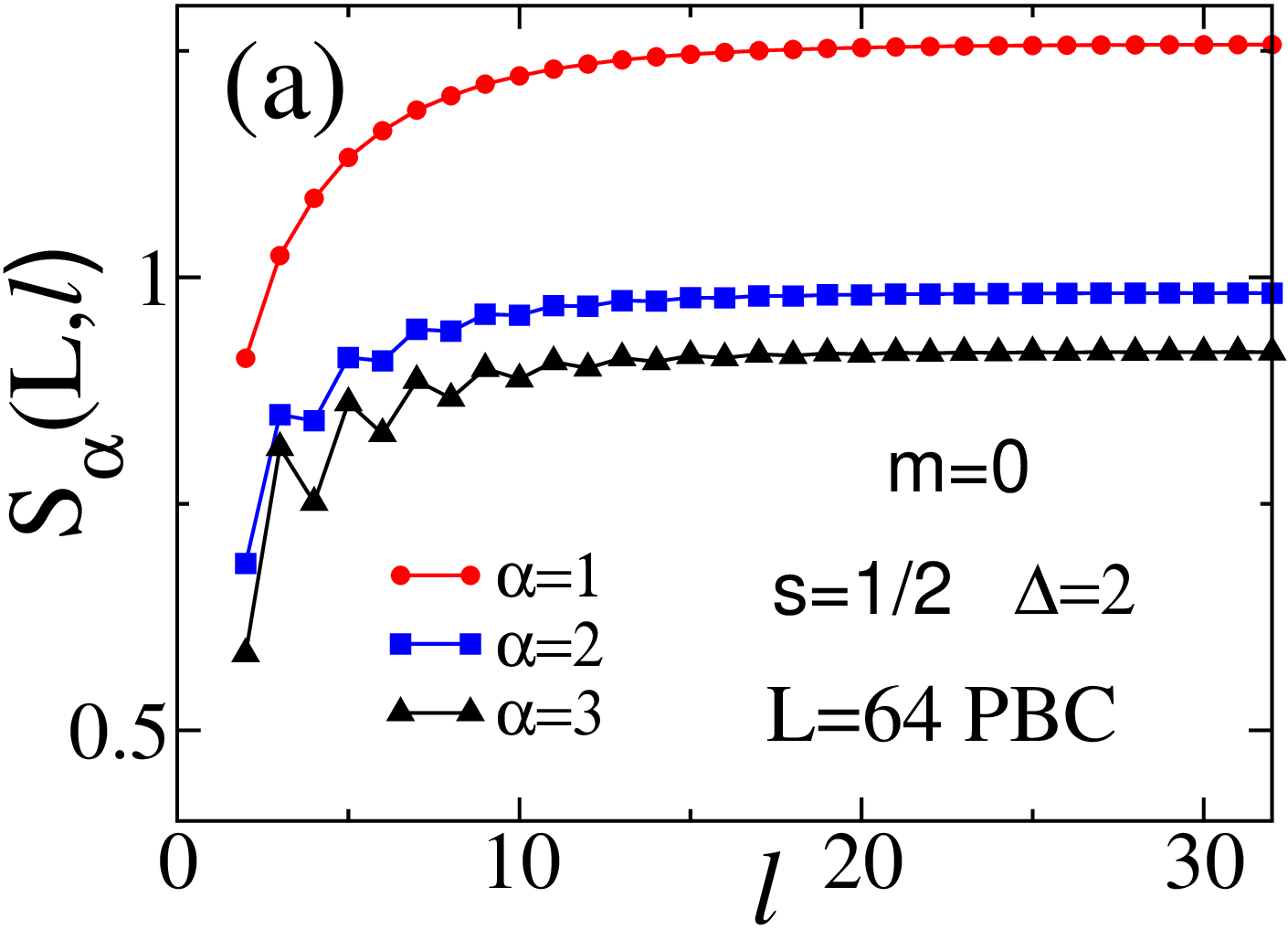} 
\par\end{centering}

\begin{centering}
\includegraphics[scale=0.27]{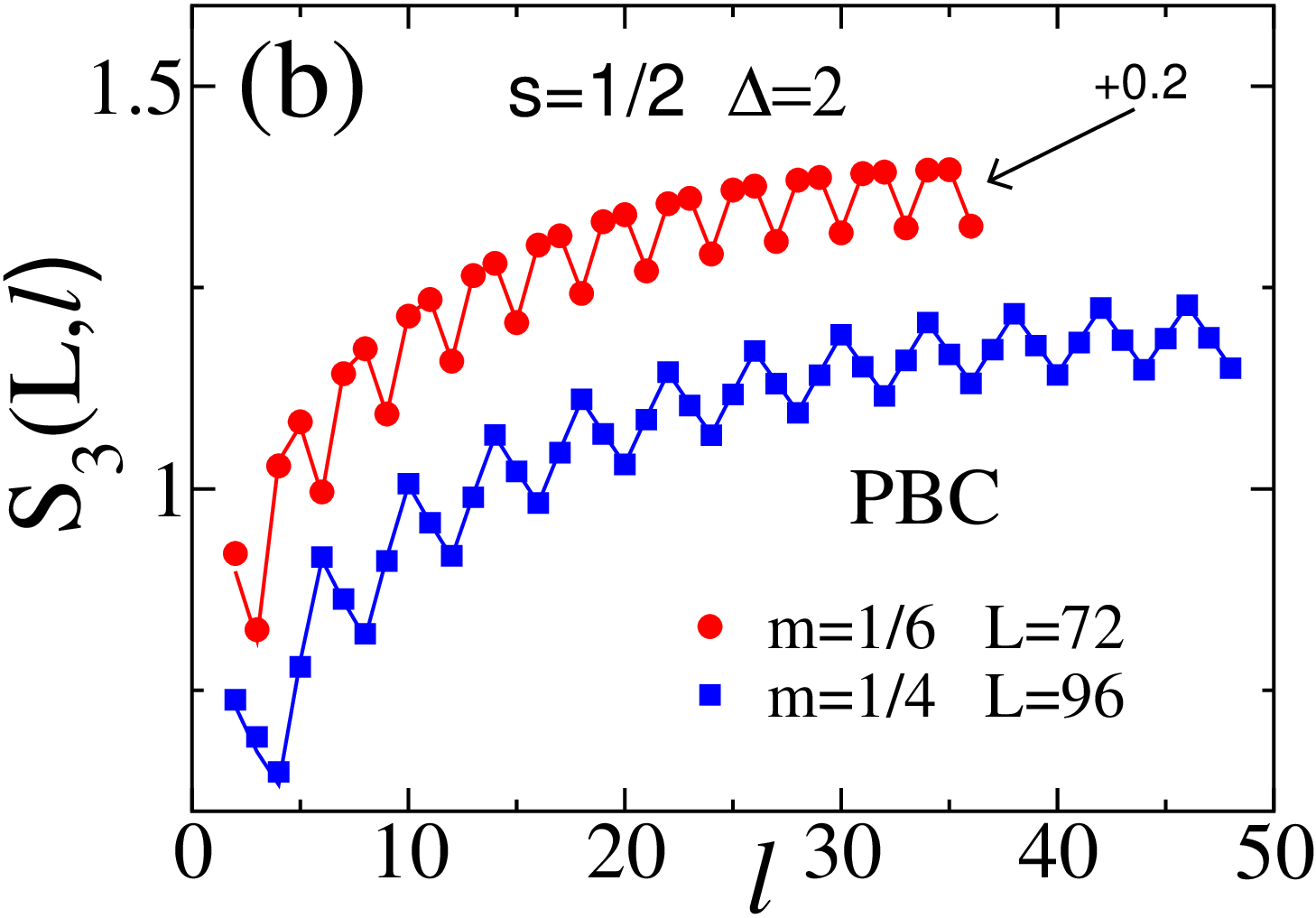} 
\par\end{centering}

\begin{centering}
\includegraphics[scale=0.27]{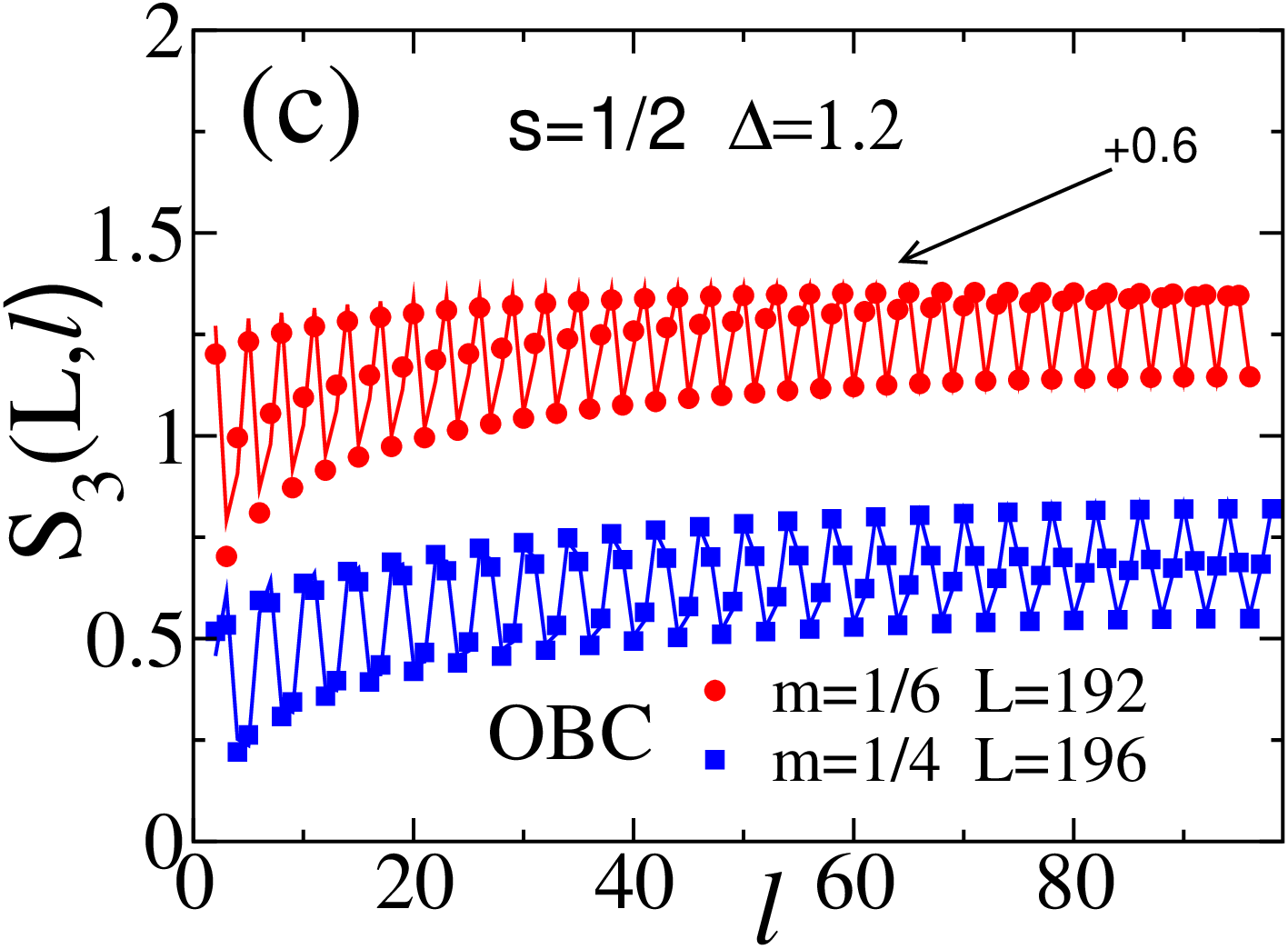} 
\par\end{centering}

\caption{\label{fig2} (Color online). The Rényi entropy $S_{\alpha}(L,l)$
as a function of $l$ for the spin-1/2 XXZ chain for $\Delta>1$.
(a) $L=96$, $\Delta=2$, PBC, and three values of $\alpha$ (see
legend). (b) $S_{3}(L,l)$ for chains with PBC, $\Delta=2$, and two
values of magnetization $m$ (see legend). (c) $S_{3}(L,l)$ for chains
with OBC, $\Delta=1.2$ and two values of $m$ (see legend).}

\end{figure}

Note, from Eqs. (\ref{eq:3c}) and (\ref{eq:3cOBC}), that the periodicity
of the oscillating term $S_{\alpha}^{osc}$ depends on the value of
the Fermi momentum $k_{F}$. On the other hand, $k_{F}$ depends of
the value of the magnetization, namely $k_{F}=(1/2-m)\pi$. Therefore,
if we change the magnetization, a change in the periodicity of the
oscillations of $S_{\alpha}$ should be observed. According to Eqs.
(\ref{eq:3c}) and (\ref{eq:3cOBC}), the period of oscillations is
$\Delta l^{*}=\pi/k_{F}$ for PBC and $\Delta l^{*}=\pi/k_{F}^{o}$
for OBC. Our results presented in Figs. 1 and 2, are in perfect agreement
with these predictions {[}see also Fig. \ref{fig3}(d){]}.

In Fig. 2, we show the Rényi entropy for $\Delta>1$ and some values
of $m$. For $|\Delta|>1$ and $m=0$ the Rényi entropies tend to
a constant,\citealp{cvidal,cardyentan} since the system is gapped.
On the other hand, for non-zero magnetization ($0<m<1/2$) the system
is critical,\citealp{xxzh1,xxzh2,xxzh3,xxzh4} therefore it is expected
that $S_{\alpha}$ behaves as Eq. (\ref{eq:3}). Indeed, we have observed
these two behaviors, as illustrated in Figs. 2(a)-(c).

The numerical values of $p_{\alpha}^{(p)}$ and $p_{\alpha}^{(o)}$
obtained from the fitting together with the corresponding predicted
exact values are presented in the Tables II and III. As we can see
in these tables, the values found are very close to the expected ones
(with the exception of $\Delta=\cos(\pi/50$)=0.9980 and $m=0$) which
strongly suggest that the amplitudes of the oscillating terms decay
as predicted and observed previously.\citealp{entropyaffleckosc,entropyosc,xxPBCh,calabreseOBC}
The fact that the estimated values of $p_{\alpha}^{(p)}$ for $\Delta=0.9980$
and $m=0$ are not so accurate, is not a surprise since, at the isotropic
point logarithmic corrections are present. Consequently the finite-size
estimates have a very slow convergence for anisotropies close to $\Delta=1$.

\begin{table}
PBC \begin{tabular}{c|c|c|c}
\hline 
 &  & $p_{2}^{(p)}$  & $p_{3}^{(p)}$\tabularnewline
\hline
\hline 
 & $m=0$  & 0.677  & 0.448\tabularnewline
 &  & (\emph{0.666})  & (\emph{0.4444})\tabularnewline
$\Delta=\sqrt{2}/2$  & $m=1/6$  & 0.757  & 0.496\tabularnewline
 &  & (\emph{0.7498})  & (\emph{0.4999})\tabularnewline
 & $m=1/4$  & 0.811  & 0.553\tabularnewline
 &  & (\emph{0.8062})  & (\emph{0.5375})\tabularnewline
\hline 
 & $m=0$  & 0.631  & 0.411\tabularnewline
 &  & (\emph{0.49})  & (\emph{0.3267})\tabularnewline
$\Delta=0.9980$  & $m=1/6$  & 0.706  & 0.462\tabularnewline
 &  & (\emph{0.6946})  & (\emph{0.4631})\tabularnewline
 & $m=1/4$  & 0.771  & 0.528\tabularnewline
 &  & (\emph{0.7659})  & (\emph{0.5106})\tabularnewline
\hline 
 & $m=1/6$  & 0.587  & 0.390\tabularnewline
$\Delta=2$  &  & \emph{(0.5905})  & (\emph{0.3937})\tabularnewline
 & $m=1/4$  & 0.675  & 0.442\tabularnewline
 &  & (\emph{0.6895})  & (\emph{0.4596})\tabularnewline
\hline 
 &  &  & \tabularnewline
\end{tabular}

\caption{The exponents $p_{2}^{(p)}$ and $p_{3}^{(o)}$ obtained through a
fit of Eq. (\ref{eq:3}) for the spin-1/2 XXZ chains with PBC, $L=96$,
and some values of $\Delta$ and $m$. We considered $c=1$ in the
fitting to Eq. (\ref{eq:3}). We also discarded the first points of
Rényi entropy $S_{\alpha}$ in the fitting procedure (up to 5). The
values in the parentheses are the exact ones. }

\end{table}

\begin{table}
OBC\begin{tabular}{c|c|c|c}
\hline 
 &  & $p_{2}^{(o)}$  & $p_{3}^{(o)}$\tabularnewline
\hline
\hline 
 & $m=0$  & 0.300  & 0.221\tabularnewline
 &  & (\emph{0.3333})  & (\emph{0.2222})\tabularnewline
$\Delta=\sqrt{2}/2$  & $m=1/6$  & 0.385  & 0.301\tabularnewline
 &  & (\emph{0.3749})  & (\emph{0.25}00)\tabularnewline
 & $m=1/4$  & 0.378  & 0.240\tabularnewline
 &  & (\emph{0.4031})  & (\emph{0.2687})\tabularnewline
 &  &  & \tabularnewline
\hline 
 & $m=1/6$  & 0.315  & 0.251\tabularnewline
$\Delta=1.2$  &  & (\emph{0.}3326)  & (\emph{0.}2217)\tabularnewline
 & $m=1/4$  & 0.339  & 0.209\tabularnewline
 &  & (\emph{0.}3722)  & (\emph{0.}248)\tabularnewline
 &  &  & \tabularnewline
\end{tabular}

\caption{Same as Table II, but for chains with OBC. We fit the data with system
sizes $L=180-196$. We have discarded the first points (up to 12 points)
in the fitting procedure. }

\end{table}

\subsection{s>1/2}

\begin{figure}
\begin{centering}
\includegraphics[scale=0.28]{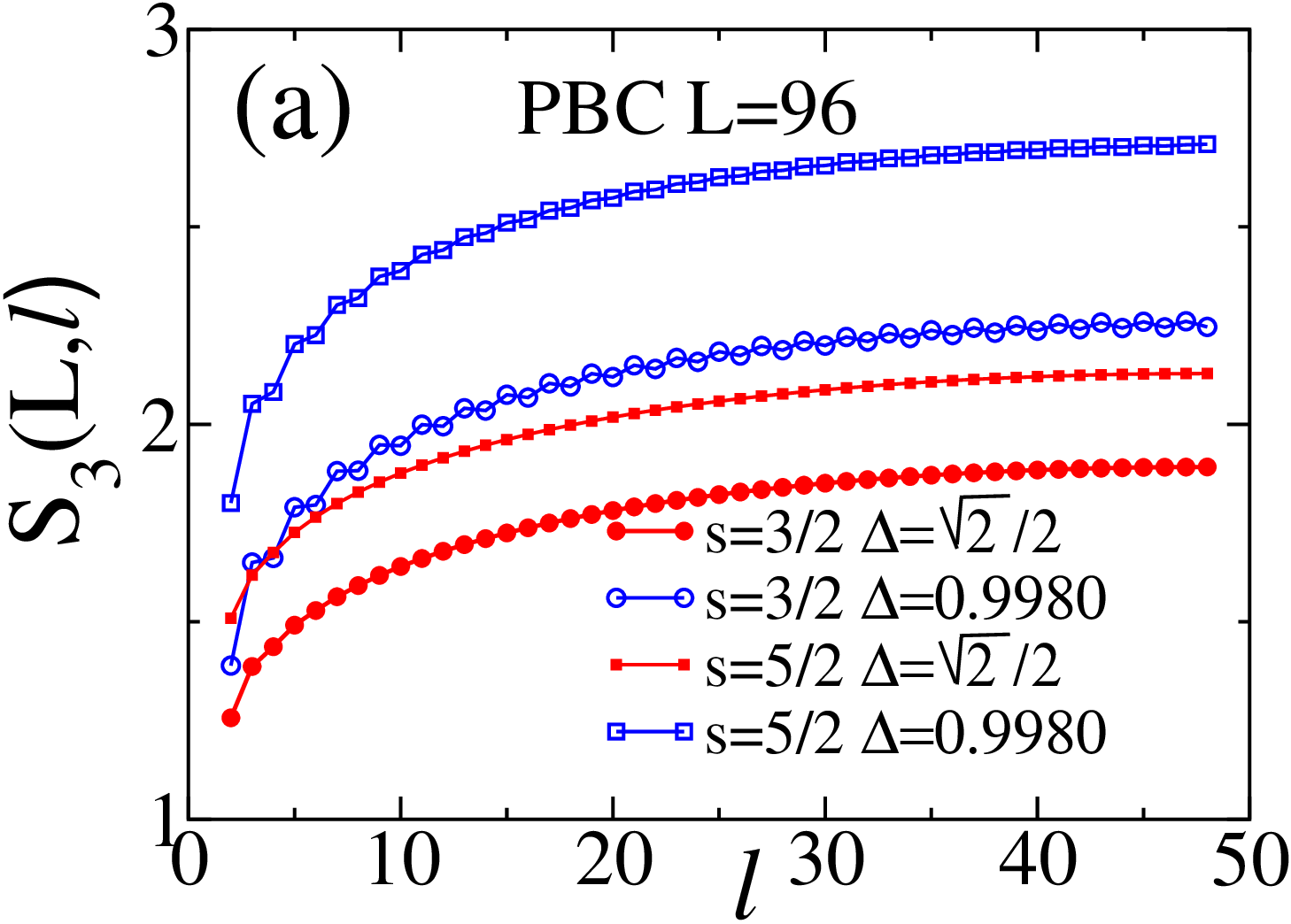} 
\par\end{centering}

\begin{centering}
\includegraphics[scale=0.28]{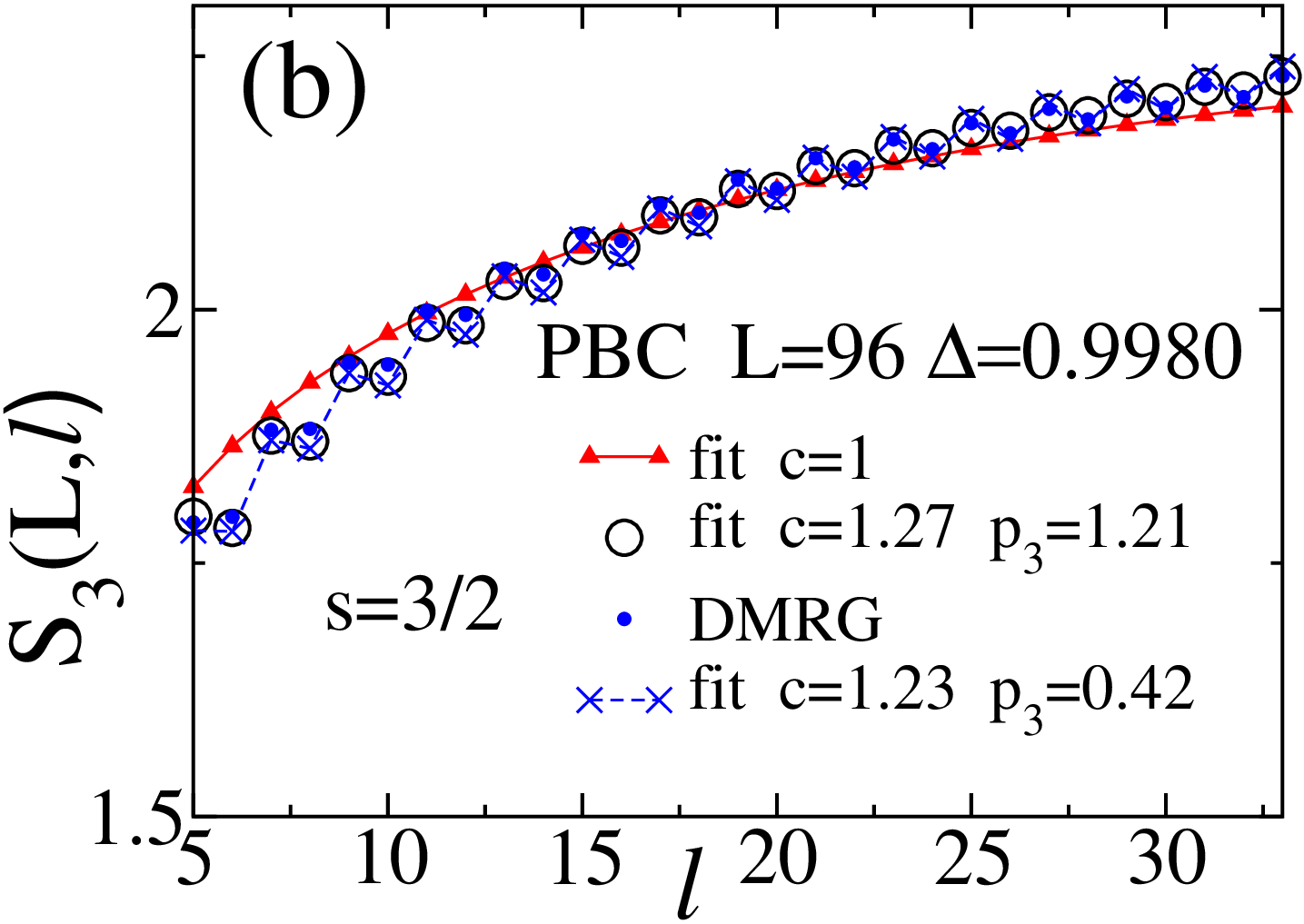} 
\par\end{centering}

\begin{centering}
\includegraphics[scale=0.28]{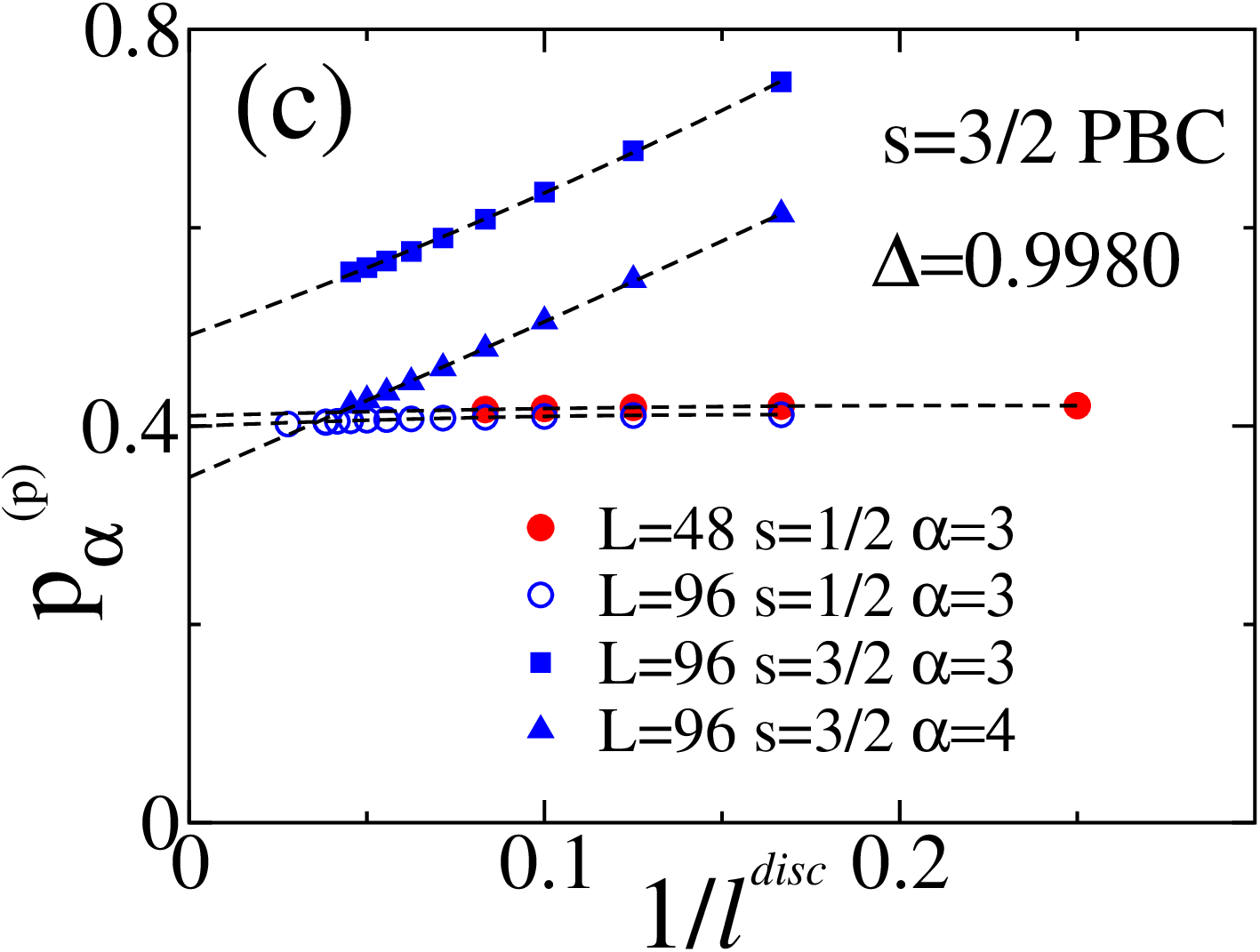} 
\par\end{centering}

\begin{centering}
\includegraphics[scale=0.28]{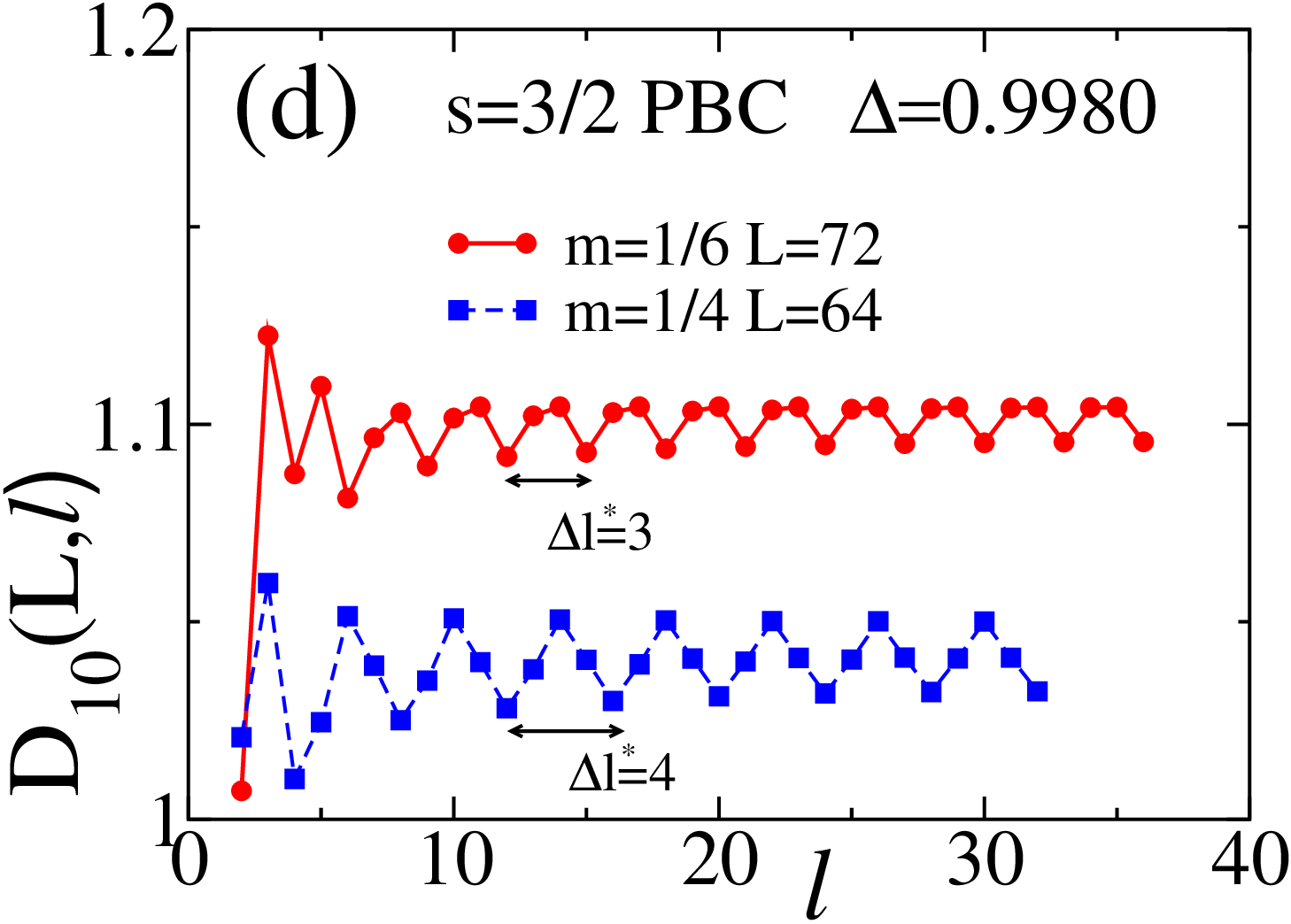} 
\par\end{centering}

\caption{\label{fig3} (Color online). (a) The Rényi entropy $S_{3}(L,l)$
vs $l$ for the spin-3/2 and 5/2 chains of size $L=96$, and two values
of $\Delta$ (see legend). (b) $S_{3}(L,l)$ vs $l$ for the spin-3/2
chain of size $L=96$ and $\Delta=0.9980$. The solid circles are
the DMRG data and the other symbols are fits (see legend). (c) Estimates
of $p_{\alpha}^{(p)}$ as function of $1/$$l^{disc}$ for the spin-1/2
and 3/2 chains with PBC. The symbols are the numerical data and the
dashed lines connect the fitted data (see text). (d) $D_{10}(L,l)$
vs $l$ for the spin-3/2 chains at $\Delta=0.9980$ and two values
of the magnetization (see legend).}

\end{figure}

Now, let us consider the case of spins $s>$1/2. As we will see below,
it is not simple to confirm that the oscillating term decays with
the exponents $p_{\alpha}^{(p)}=2p_{\alpha}^{(o)}=\frac{2K}{\alpha}$.
The difficulty of extracting $p_{\alpha}^{(p)}$ and $p_{\alpha}^{(o)}$
is mainly due to the small amplitude of the oscillations of $S_{\alpha}^{osc}$,
as we are going to see below. In order to get some idea of their order
of the magnitude, let us consider the oscillations in the region around
the middle of the chains. According to Eqs. (\ref{eq:3c}) and (\ref{eq:3cOBC})
the amplitudes are $A_{s}^{(p)}\sim g_{\alpha}^{(p)}(\Delta,s)/L^{p_{\alpha}^{(p)}(\Delta,s)}$
and $A_{s}^{(o)}\sim g_{\alpha}^{(o)}(\Delta,s)/L^{p_{\alpha}^{(0)}(\Delta,s)}$.
However, our numerical results indicate that $g_{\alpha}^{(p)}$ and
$g_{\alpha}^{(o)}$are practically independent of the anisotropy value.
The amplitudes $A_{s}^{(p)}$ and $A_{s}^{(o)}$ are them expected
to be smaller as $p_{\alpha}^{(p)}$ and $p_{\alpha}^{(o)}$ decrease
or, equivalently, as $x_{p}$ increases. For example, for $\Delta=\sqrt{2}/2$
($\gamma=\pi/4$), the exponents are $x_{p}=0.333$, 0.099, and $0.057$
for $s=1/2,$ $3/2$ and 5/2, respectively.\citealp{xavierentanglement}
Therefore, we expect, at this anisotropy, a decreasing of the amplitude
of the oscillations as $s$ increases. This can be observed when we
compare Fig. \ref{fig1}(a) with the curves of Fig. \ref{fig3}(a)
at $\Delta=\sqrt{2}/2$. Actually, our results indicate that for $s=3/2$
and $s=5/2$ the amplitudes of the oscillations are approximately
one and two order of magnitude smaller, as compared with those of
the $s=1/2$ at $\Delta=\sqrt{2}/2$, respectively. With such small
amplitudes it is quite difficult to extract, at the anisotropy $\Delta=\sqrt{2}/2$,
accurate values of $p_{\alpha}^{(p)}$ and $p_{\alpha}^{(o)}$ for
the spin chains with $s>1/2$. 

In order to estimate the exponents $p_{\alpha}^{(p)}$ and $p_{\alpha}^{(o)}$
with some accuracy for quantum spin chains with $s>$1/2, we need
to find at least a region in the critical phase of the model where
$x_{p}\sim0.3$, as in the spin-1/2 case. As it is well known, at
the isotropic point $\Delta=1$ all spin-$s$ Heisenberg chains have
$x_{p}=1/2$. However, at this isotropic point, the operator governing
the finite-size corrections is marginal producing logarithmic corrections
that make a slow convergence in the finite-size estimates. We then
select an anisotropy close to the isotropic point. We choose $\Delta=\cos\pi/50=0.9980$,
although we should expect that even at this point the finite-size
corrections are very large producing only rough estimates of $x_{p}$,
$p_{\alpha}^{(p)}$ and $p_{\alpha}^{(o)}$.

In Table IV, we present the anomalous dimensions $x_{p}$ for the
spin-3/2 and 5/2 chains for $\Delta=0.9980$ and some values of magnetization
$m$. Note that $x_{p}$ <0.15, for the spin-5/2 quantum chains. This
means that the evaluation of $p_{\alpha}^{(p)}$ and $p_{\alpha}^{(o)}$
will be quite difficult even taking anisotropies close to the isotropic
point. For this reason we concentrate in the $s=3/2$ case at $\Delta=0.9980$
and $m=0$, where the estimate value is $x_{p}=0.39$. This value
is close to the value $x_{p}=0.375$ (see Table I) for the spin-1/2
with $\Delta=\sqrt{2}/2$. We then, naively, expect that the amplitudes
$A_{s}^{(p)}$ and $A_{s}^{(o)}$ are of the same order in these two
cases. However, to our surprise the amplitudes of the oscillations
for $s=3/2$ are quite smaller than those of the $s=1/2$, as we can
see in Figs. \ref{fig3}(a) and \ref{fig3}(b). Similar results were
also observed for the $s=3/2$ chains with OBC and $m\ne0$ (for $\alpha=1$
and $m=0$, see also Ref. \onlinecite{xavierentanglement}). These
results indicate that the non-universal constants $g_{\alpha}^{(p)}$
and $g_{\alpha}^{(o)}$ appearing in Eqs. (\ref{eq:3c}) and (\ref{eq:3cOBC})
are very small for $s>$1/2, as compared with the $s=1/2$ case. This
means that the evaluation of $p_{\alpha}^{(p)}$ and $p_{\alpha}^{(o)}$
directly from the decaying of the Rényi entropy oscillation, for the
$s>$1/2 quantum chains, is a quite hard task.

On despite of the above difficulties, we try to estimate $p_{\alpha}^{(p)}$
and $p_{\alpha}^{(o)}$ for the spin-3/2 chain at the anisotropy $\Delta=0.9980$
and magnetization $m=0$. As already mentioned, we should expect rough
estimates since we are close the the isotropic point. Like the $s=1/2$
case the oscillations, although smaller, happens only for $\alpha>1$.
The central charge obtained by fitting Eq. (\ref{eq:3}) is, in the
$\alpha=1$ case, $c=1.01$ at $\Delta=\sqrt{2}/2$ and $c=1.2$ at
$\Delta=0.9980$. If we fix $c=1$ the fitting is quite poor at $\Delta=0.9980$.
A nice fit, at this anisotropy, is only obtained by allowing $c$
as a free parameter. Certainly the estimate of $c,$ as we increase
$L$, decreases towards the exact value $c=1$. Let us now concentrate
on the case of $\alpha=3$ where the amplitudes are clearly present.
We are going to fit the data with Eq. (\ref{eq:3}) in three distinct
ways. 

Firstly, we try to fit the DMRG data to Eq. (\ref{eq:3}) by considering
$c=1$ fixed. In this case, the least square fitting give us $g_{3}^{(p)}=0$
for the best fit with $\chi^{2}=0.0426$\cite{comentario2} {[}red
triangles in Fig. \ref{fig3}(b){]}. This means that through this
procedure we get no oscillations\cite{comentario} {[}see Fig. \ref{fig3}(b){]}
and we are not able to extract values of $p_{3}^{(p)}$. 

\begin{table}
\begin{tabular}{ccccc}
\hline 
 & $m=0$  & $m=1/6$  & $m=1/4$  & $m=3/10$\tabularnewline
\hline
\hline 
$s=3/2$  & 0.39  & 0.1235  & 0.1177  & 0.1157\tabularnewline
$s=5/2$  & 0.12  & 0.074  & 0.071  & 0.070\tabularnewline
\hline
\end{tabular}

\caption{The anomalous dimension $x_{p}$ for the anisotropic spin-3/2 and
5/2 Heisenberg chains with PBC, with anisotropy $\Delta=0.9980$ and
some values of the magnetization $m$. }

\end{table}

In the second procedure we allow $c$ and $p_{3}^{(p)}$ as free fit
parameters. In this case we get $p_{3}=1.21$ and $c=1.27$ with $\chi^{2}=0.0041$
for $L=96$ {[}black cicles in Fig. \ref{fig3}(b){]}. Similar results
were obtained for $L=48$ where we get $p_{3}^{(p)}=1.19$. 

In the third procedure we fix the expected value of $p_{3}^{(p)}=\frac{1}{6x_{p}}=1/(6\times0.39)=0.427$.
In this case the get $\chi^{2}=0.0087$ which is higher than previous
procedure. In Fig. \ref{fig3}(b), the blue dashed line connects the
fit to our data by considering the predicted exponent $p_{3}^{(p)}$=0.427
fixed. Although the finite-size corrections are large, since we are
close to the isotropic point, the results suggest that the oscillating
term of the Rényi entropy decays as Eq. (\ref{eq:3c}). We note that
for the spin-3/2 chains, differently from the spin-1/2 case, the estimates
of $p_{\alpha}^{(p)}$  depend strongly on how many sites $l$ of
the subsystem $A$ we discard. In Fig.\ref{fig3}(c), we present the
estimates of the exponents $p_{3}^{(p)}$ and $p_{4}^{(p)}$ as function
of $1/$$l^{disc}$, where $l^{disc}$ is the number of sites we discard
in the fit procedure (we discard the sites $l=1,\ldots,l^{disc}$).
As we can note in this figure, for the spin-1/2 case the estimates
of $p_{\alpha}^{(p)}$ weakly depend on the values $l^{disc}$ and
$L$. The weak dependence with the lattice size $L$ also happens
for $s=3/2$. However, contrarily to the spin-1/2 case, those estimates
are very sensitive to the number of sites $l^{disc}$ of the subsystem
$A$ we discard, as shown in Fig.\ref{fig3}(c). In order to take
into account this effect, we assume that $p_{\alpha}^{(p)}(l^{disc})$
behaves as

\begin{equation}
p_{\alpha}^{(p)}(l_{\alpha}^{disc})=p_{\alpha}^{(p)}+\frac{a_{\alpha}}{l^{disc}}+\frac{b_{\alpha}}{(l^{disc})^{2}}.\label{eq:palpha}\end{equation}
 If we fit our data for the spin-3/2 case {[}presented in Fig. \ref{fig3}(c){]}
with this equation, we obtain $p_{3}^{(p)}=0.49$ and $p_{4}^{(p)}=0.34$.
These values are quite close to the predicted ones ($p_{3}^{(p)}=0.427$
and $p_{4}^{(p)}=0.32$). In the above fit procedure the central charge
$c$ is also a free fit parameter. We also observed, in this case,
that the estimates of $c$ are sensitive to the values of $l^{disc}$,
and as we increase $l^{disc}$ they get closer to the expected value
$c=1$. In Fig.\ref{fig3}(c), we present the fits for the spin-3/2
chains only for $l^{disc}<20$, since for large values of $l$ the
amplitudes of the oscillations are of the same order of the numerical
errors. Finding estimates of the exponent $p_{\alpha}^{(o)}$ for
the spin-3/2 chains with OBC are even more difficult. In this case
the estimates of $p_{\alpha}$ are very sensitive to $l^{disc}$ \emph{and
also} to $L$. For this reason we are not able to find a simple procedure
to estimate this exponent for the spin-3/2 chains with OBC.

We could naively expect that for $m>0$, where logarithmic corrections
are not expected, $p_{\alpha}^{(p)}$ would be better estimated. However,
for $m>$0 the anomalous dimension $x_{p}$ is small (see Table IV)
and consequently the amplitudes of the oscillations are also small,
complicating our analysis, as discussed earlier. However, we can observe
an important feature of Eq. (\ref{eq:3c}) in the Rényi entropy of
the spin-3/2 chains. To better see this feature, it is convenient
to define the difference $D_{\alpha}(L,l)=S_{\alpha}-S_{\alpha}^{CFT}$.
In Fig. \ref{fig3}(d), we present $D_{10}(L,l)$ as a function of
$l$ for the spin-3/2 chain for $\Delta=0.9980$ and two values of
the magnetization. We choose to present a large value of $\alpha$
since the amplitudes are bigger as we increase $\alpha.$ According
to Eq. (\ref{eq:3c}) the period of the oscillations is $\Delta l^{*}=\pi/k_{F}=2/(2-2m)$.
In fact, we have observed this periodicity, as shown in Fig. \ref{fig3}(d)
for two values of the magnetization.

\section{DISCUSSION}

In this paper, we investigate the Rényi entropies of the spin-$s$
anisotropic Heisenberg chains in a magnetic field. These quantum chains
are critical and conformal invariant in a wide region of values of
the magnetic field $h$ and anisotropy $\Delta$. The long-distance
physics of this critical region is described by a Luttinger Liquid
CFT, with central charge $c=1$. For this reason, these models are
very attractive for testing predictions for one-dimensional critical
systems. In particular, it is expected that the $\alpha-$Rényi entropies
have a term that oscillates with the subsystem size, whose amplitudes
show a power-law decay with universal exponents $p_{\alpha}^{(p)}$
and $p_{\alpha}^{(o)}$. These exponents are expected to depend on
the Luttinger parameter $K$, \emph{i. e.,} $p_{\alpha}^{(p)}=2K/\alpha$
and $p_{\alpha}^{(o)}=K/\alpha$ for PBC and OBC, respectively. This
universal behavior was obtained exactly for the spin-1/2 XXZ chains
with a magnetic field with PBC and OBC for $\Delta=0$ (XX chains).\citealp{xxPBCh,calabreseOBC}
Moreover, for $h=0$, DMRG calculations of the spin-1/2 XXZ chain
also indicate that the oscillating term indeed decays as predicted.\citealp{entropyosc}
As part of this work, we made an extensive study of the spin-1/2 model,
but considered a much wider region of couplings than those considered
earlier. Using DMRG technique, we also investigate extensively the
quantum spin-$s$ chains (up to $s=5/2)$ with PBC/OBC for several
values of anisotropy $\Delta$ and magnetic field. Using the CFT machinery
we were able to get accurate estimates of the Luttinger parameter
$K$. For the spin-1/2 chains with PBC and OBC, we extract the exponents
$p_{\alpha}^{(p)}$ and $p_{\alpha}^{(o)}$ through a fit of Eq. (\ref{eq:3})
and confirm the predicted universal behavior of the Rényi entropy
for several values of magnetization $m$ and anisotropy $\Delta$.
For spin $s=3/2$ our estimates of the the exponent $p_{\alpha}$
are not so accurate due to the fact that the non-universal constants
$g_{\alpha}^{(p)}$ and $g_{\alpha}^{(o)}$ are very small (typically
one order of magnitude smaller than the ones of spin-1/2 chain). Even
though, our results indicate that $p_{\alpha}^{(p)}$ and $p_{\alpha}^{(o)}$
are related with the Luttinger parameter, as predicted. We also observe
that the periodicity of the oscillating term changes with the magnetization,
as conjectured. For $s>3/2$ we were not able to extract the exponents
$p_{\alpha}^{(p)}$ and $p_{\alpha}^{(o)}$, mainly due to the fact
that the $g_{\alpha}^{(p)}$ and $g_{\alpha}^{(o)}$ are so small,
making the data difficult to analyze. In this case, the Luttinger
parameter $K$ is larger (compared with the ones of the spin-1/2)
and also contributes to make the oscillating term $S^{osc}$ almost
imperceptible. We could even think that this term is null as happens
in the Ising model,\citealp{igloixy,xxPBCh} and that such small oscillations
comes from numerical errors. However we are convinced that those small
oscillations are not related with the truncation errors in the DMRG.
We have observed that those oscillations do not decrease as we increase
the number states kept (up to $\tilde{m}=4000$) in the DMRG procedure.
These results strongly indicate that the non-universal amplitudes
$g_{\alpha}^{(p)}$ and $g_{\alpha}^{(o)}$ decrease very fast as
the spin $s$ increases. This makes a huge challenge the determination
of the exponents $p_{\alpha}^{(p)}$ and $p_{\alpha}^{(o)}$ with
reasonable accuracy for $s>$1/2. 
\begin{acknowledgments}
The authors are grateful to P. Calabrese and J. Hoyos for a careful
reading of the manuscript and helpful discussions. This Research supported
by the Brazilian agencies FAPEMIG, FAPESP, and CNPq. 
\end{acknowledgments}

\bibliographystyle{apsrev4-1}
\addcontentsline{toc}{section}{\refname}

\begin{thebibliography}{10}
\bibitem{comentario2}%
  \BibitemOpen
    \bibinfo {note} {For a set of N data $d_i$
  $(i=1,...N)$ whose fitted points are $f_i$ the $\chi^2$ is given by 
$\left( \sum_i^N(f_i-d_i)^2 \right)^{1/2}$ }%
  \bibAnnoteFile{NoStop}{comentario2}%
\bibitem{comentario}%
  \BibitemOpen
    \bibinfo {note} {If we discard the first sites,
  the oscillations are "recovered". However, if we consider $c$ as a free
  parameter, we obtain a better fit.}%
  \bibAnnoteFile{Stop}{comentario}%
\end{thebibliography}
%

\end{document}